\documentclass[sn-nature]{sn-jnl}

\usepackage{graphicx}%
\usepackage{multirow}%
\usepackage{amsmath,amssymb,amsfonts}%
\usepackage{amsthm}%
\usepackage{mathrsfs}%
\usepackage[title]{appendix}%
\usepackage{xcolor}%
\usepackage{textcomp}%
\usepackage{manyfoot}%
\usepackage{booktabs}%
\usepackage{algorithm}%
\usepackage{algorithmicx}%
\usepackage{algpseudocode}%
\usepackage{listings}%
\usepackage{placeins}
\usepackage{lineno}

\usepackage{geometry}
\begin{document}

\title[Detectorless 3D terahertz imaging: achieving subwavelength resolution with reflectance confocal interferometric microscopy]{Detectorless 3D terahertz imaging: achieving subwavelength resolution with reflectance confocal interferometric microscopy}





\author[1]{\fnm{Jorge} \sur{Silva}}
\author[1]{\fnm{Martin} \sur{Pl\"{o}schner}}
\author[1]{\fnm{Karl} \sur{Bertling}}
\author[1]{\fnm{Mukund} \sur{Ghantala}}
\author[1]{\fnm{Tim} \sur{Gillespie}}
\author[1]{\fnm{Jari} \sur{Torniainen}}
\author[1]{\fnm{Jeremy} \sur{Herbert}}
\author[1]{\fnm{Yah Leng} \sur{Lim}}
\author[2]{\fnm{Thomas} \sur{Taimre}}
\author[1]{\fnm{Xiaoqiong} \sur{Qi}}
\author[1]{\fnm{Bogdan C.} \sur{Donose}}
\author[1]{\fnm{Tao} \sur{Zhou}}
\author[1]{\fnm{Hoi-Shun} \sur{Lui}}
\author[3]{\fnm{Dragan} \sur{Indjin}}
\author[3]{\fnm{Yingjun} \sur{Han}}
\author[3]{\fnm{Lianhe} \sur{Li}}
\author[3]{\fnm{Alexander} \sur{Valavanis}}
\author[3]{\fnm{Edmund H.} \sur{Linfield}}
\author[3]{\fnm{A. Giles} \sur{Davies}}
\author[3]{\fnm{Paul} \sur{Dean}}
\author*[1]{\fnm{Aleksandar D.}\sur{Raki\'{c}}}\email{a.rakic@uq.edu.au}

\affil[1]{\orgdiv{School of Electrical Engineering and Computer Science}, \orgname{The University of Queensland}, \orgaddress{\city{Brisbane}, \postcode{4072}, \state{Queensland}, \country{Australia}}}

\affil[2]{\orgdiv{School of Mathematics and Physics}, \orgname{The University of Queensland}, \orgaddress{\city{Brisbane}, \postcode{4072}, \state{Queensland}, \country{Australia}}}

\affil[3]{\orgdiv{School of Electronic and Electrical Engineering}, \orgname{University of Leeds}, \orgaddress{\city{Leeds}, \postcode{LS2 9JT}, \country{UK}}}


\abstract{
Terahertz imaging holds great potential for non-destructive material inspection, but practical implementation has been limited by resolution constraints. In this study, we present a single-pixel THz imaging system based on a confocal microscope architecture, utilising a quantum cascade laser as both transmitter and phase-sensitive receiver. Our approach integrates laser feedback interferometry detection to achieve a two-fold improvement in lateral resolution and a two-order-of-magnitude enhancement in axial resolution over conventional imaging through precise interferometric phase measurements. This translates to a lateral resolution near $\lambda/2$ and a depth of focus better than $\lambda/5$, significantly outperforming traditional confocal systems. The system can produce a 0.5 Mpixel image in under two minutes, surpassing both raster-scanning single-pixel and multipixel focal-plane array-based imagers. Coherent operation enables simultaneous amplitude and phase image acquisition, and a custom visualisation method links amplitude to image saturation and phase to hue, enhancing material characterisation. A 3D tomographic analysis of a silicon chip reveals subwavelength features, demonstrating the system's potential for high-resolution THz imaging and material analysis. This work sets a new benchmark for THz imaging, overcoming key challenges and opening up transformative possibilities for non-destructive material inspection and characterisation.
}

\newgeometry{a4paper,         
  textwidth=13.5cm,  
  textheight=25cm, 
  heightrounded,   
  hratio=1:1,      
  vratio=2:3,
  }
\maketitle
\restoregeometry 

\section*{Introduction}\label{sec1}
Terahertz (THz) radiation plays a crucial role in imaging hard-to-characterise features due to its ability to penetrate materials that are opaque to other parts of the electromagnetic spectrum$\,$\cite{mittleman2018twenty, guerboukha2018toward, cooper2011thz, koch2015reflection, novikova2018non}. Its non-ionising nature, low phototoxicity, and capacity to identify biological molecules through their unique "fingerprints" at THz frequencies make it an emerging tool in biosensing and medical diagnostics$\,$\cite{son2019potential, markelz2022perspective, qi2023terahertz, qi2024terahertz}. Additionally, THz imaging has been applied to study carrier dynamics in silicon wafers, organic solar cells, and for conductivity mapping of novel materials like graphene$\,$\cite{ulbricht2011carrier, mezzapesa2014imaging, lane2015hot, buron2012graphene}. This rapidly expanding application space of THz imaging creates a burgeoning need for more compact technology with higher spatial and temporal resolution, as well as enhanced material specificity. However, current THz technology struggles to meet all these demands simultaneously. 

State-of-the-art THz multipixel cameras$\,$\cite{al20121,oda2015microbolometer, sizov2018terahertz} capture only the intensity of THz radiation, neglecting the phase information essential for recovering material properties like the complex refractive index. While digital holography approaches$\,$\cite{locatelli2015real, petrov2016application, yamagiwa2018real} can be used in conjunction with THz cameras to retrieve the phase information, the process is slow, computationally challenging, and suffers from phase instabilities due to the need for a reference beam. Single-pixel detection schemes, especially those using time-domain spectroscopy (TDS)$\,$\cite{jepsen2011terahertz, neu2018tutorial, koch2023terahertz}, overcome many of the limitations of multipixel cameras but still suffer from low spatial resolution and slow image formation.

Here, we demonstrate a single-pixel camera based on a terahertz confocal microscope capable of high-resolution, high-frame-rate far-field THz imaging with a unique capacity for depth sectioning and a depth resolution exceeding $\lambda/5$. The system achieves compactness and ease of alignment through its incredibly simple design: a single laser used for both illumination and detection via laser feedback interferometry (LFI)$\,$\cite{rakic2013swept}. 
Terahertz radiation from a quantum cascade laser (QCL)$\,$\cite{kohler2002terahertz, williams2007terahertz, Faist2013} is focused onto the sample for raster scanning. 
The reflected signal then travels back through the optical system and is reinjected into the QCL, which simultaneously functions as the detector, confocal pinhole, and illumination source. Combining these functions into a single device not only enables a compact design but also significantly simplifies system alignment$\,$\cite{rakic2019sensing}.
We also provide, to the best of our knowledge, the first comprehensive guide on how to implement and optimise this technology$\,$\cite{dean2011terahertz, dean2013coherent, keeley2015three, wienold2016real, pogna2021terahertz} to achieve optimal imaging performance. 

Through simulation of the beam propagation, we find the most favorable detection plane to be the QCL’s front facet. The system's point spread function (PSF), measured using a knife-edge technique, shows a better than two-fold improvement in lateral and axial confocal resolution. We attribute this enhanced resolution to the confocal imaging configuration and the characteristics of the QCL which acts as a confocal aperture. The depth of focus of $\lambda/5$, acquired from the phase information, exceeds  limitations of confocal imaging systems, resulting in a further significant improvement in axial resolution. We combine the high-resolution capability with a fast beam steering configuration to deliver a 0.5 Mpixel image in less than two minutes. This compares favorably with other reported imagers of both the raster scanning (single pixel) and multipixel focal-plane array based solutions, alike. To accommodate a wide range of applications requiring different scan areas, we also investigate the system's performance with different focusing objective numerical apertures. Finally, we illustrate the system's depth of focus resolution by performing 3D tomographic imaging of hidden features of a packaged silicon chip.

\section*{Results}\label{sec2}

\subsection*{Operating principle}
\begin{figure}[!b]
\centering\includegraphics[width=1.\textwidth]{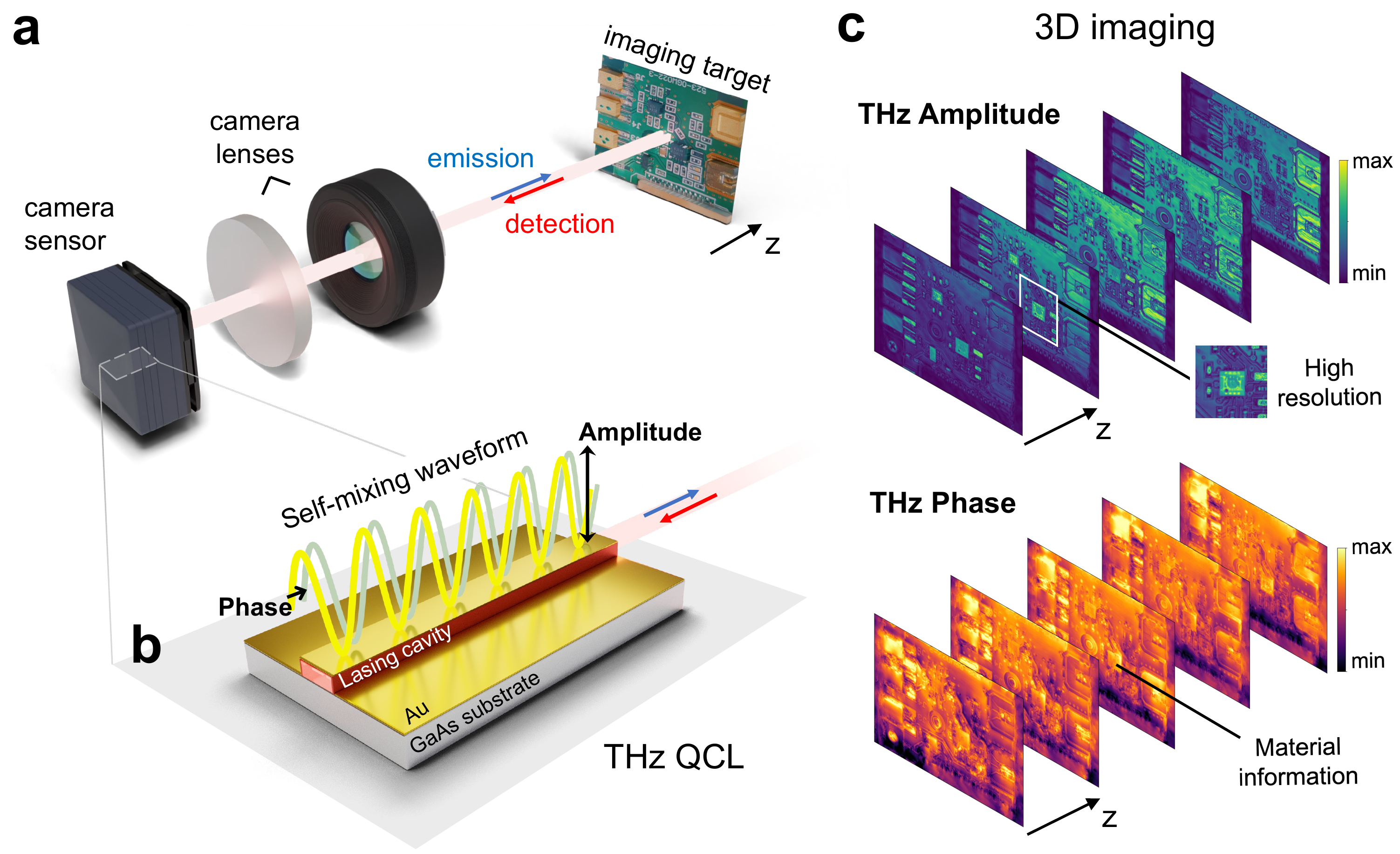}
\caption{\textbf{Camera-like 3D coherent imaging.} \textbf{a} A camera-like imaging configuration is shown, where the laser source doubles as a sensor. The optical system collimates the laser beam and focuses it onto the sample. By adjusting the numerical aperture of the focusing lens, both the spot size and depth of focus are controlled. \textbf{b} The terahertz quantum cascade laser acts as the source for the self-mixing waveform. The self-mixing effect enables coherent imaging by providing both amplitude and phase information, forming the basis for high-resolution 3D imaging. \textbf{c} Fast beam steering enables rapid image acquisition and retrieval of 3D information$\,$\cite{rakic2019sensing}. This technique allows for detailed inspection of internal structures, such as in the 3D imaging of a printed circuit board.}
\label{fig:concept}
\end{figure}

Imaging is achieved by utilising a detection scheme built into the THz source itself, along with transmission optics to capture the light reflected from the sample back into the detection sensor, similar to a camera lens (Fig. \ref{fig:concept}a). The detection principle, based on laser feedback interferometry, measures the interference between the QCL intracavity electric field and the reinjected radiation reflected from the sample$\,$\cite{dean2023}, as shown in Fig. \ref{fig:concept}b. This process relies on the self-mixing (SM) effect, which causes changes in the laser voltage, allowing the acquisition of both amplitude and phase information$\,$\cite{rakic2013swept, lim2019coherent}. Coherent 3D imaging of the sample is achieved through fast beam steering (Fig. \ref{fig:concept}c).
\clearpage
\subsection*{System characterisation and performance analysis}\label{sec:qcl_emission}
\begin{figure}[!b]
\makebox[\textwidth][c]{\includegraphics[width=\textwidth]{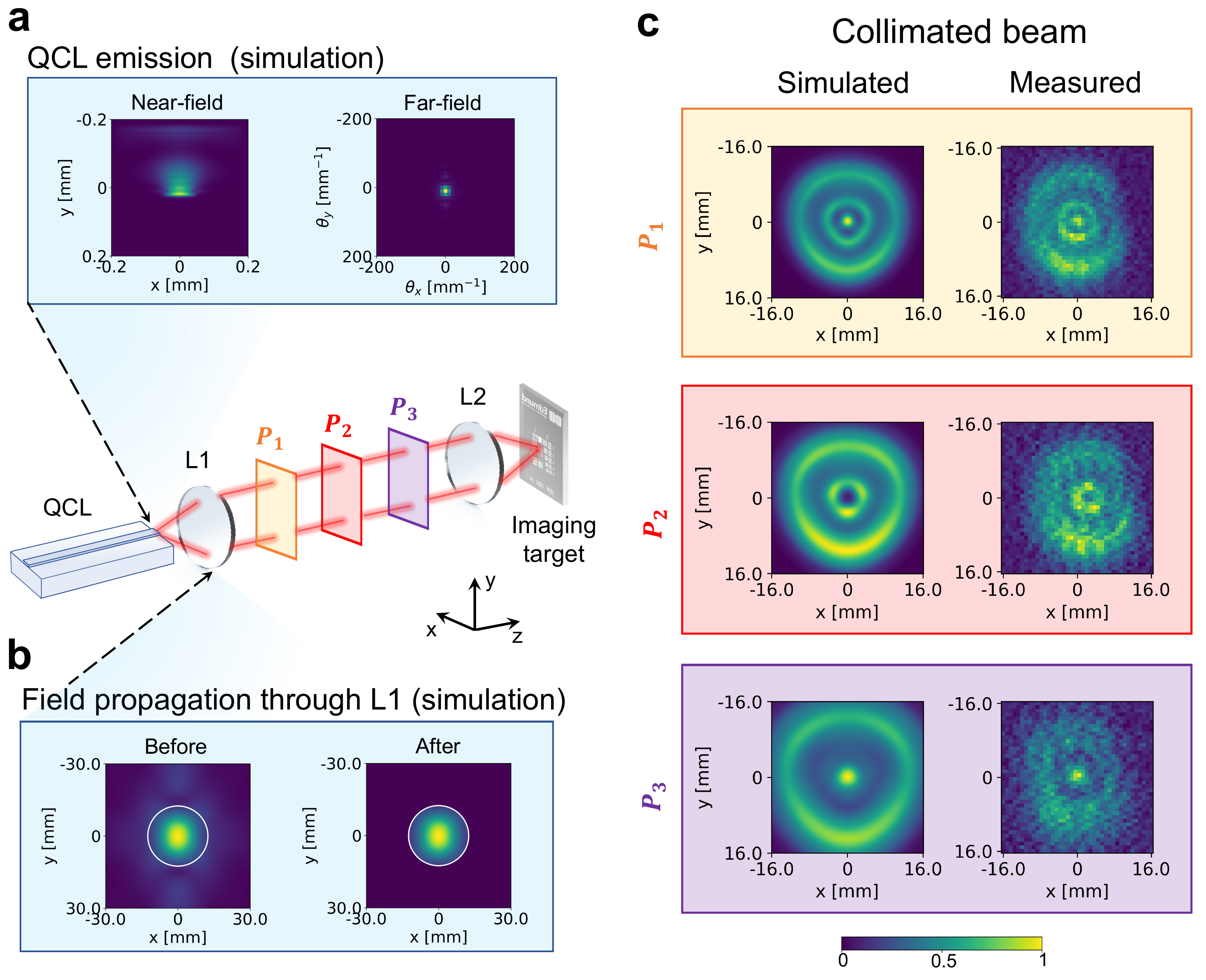}}
\caption{\textbf{QCL emission and collimation.} \textbf{a} Simulated near-field and far-field QCL emission intensity profiles. The near-field profile corresponds to the QCL's front facet emission, for which we implement a FFT to compute the far-field. \textbf{b} Simulated field propagation through a 30 mm focal length lens L1 (Tsurupica-RR-CX-1.5-30-SPS, BBLaser Inc., Tokyo, Japan), where the white circle denotes the clear aperture of the lens ($25$ mm). \textbf{c} Simulated and measured collimated beam intensity profiles for three distinct planes located after the collimating lens L1, at a distance of 480 mm ($P_1$), 630 mm ($P_2$) and 930 mm ($P_3$) from the QCL's front facet. The measured profiles were obtained by performing a 2D scan of the pyroelectric detector.}
\label{fig3}
\end{figure}

\begin{figure}[!b]
\makebox[\textwidth][c]{\includegraphics[width=\textwidth]{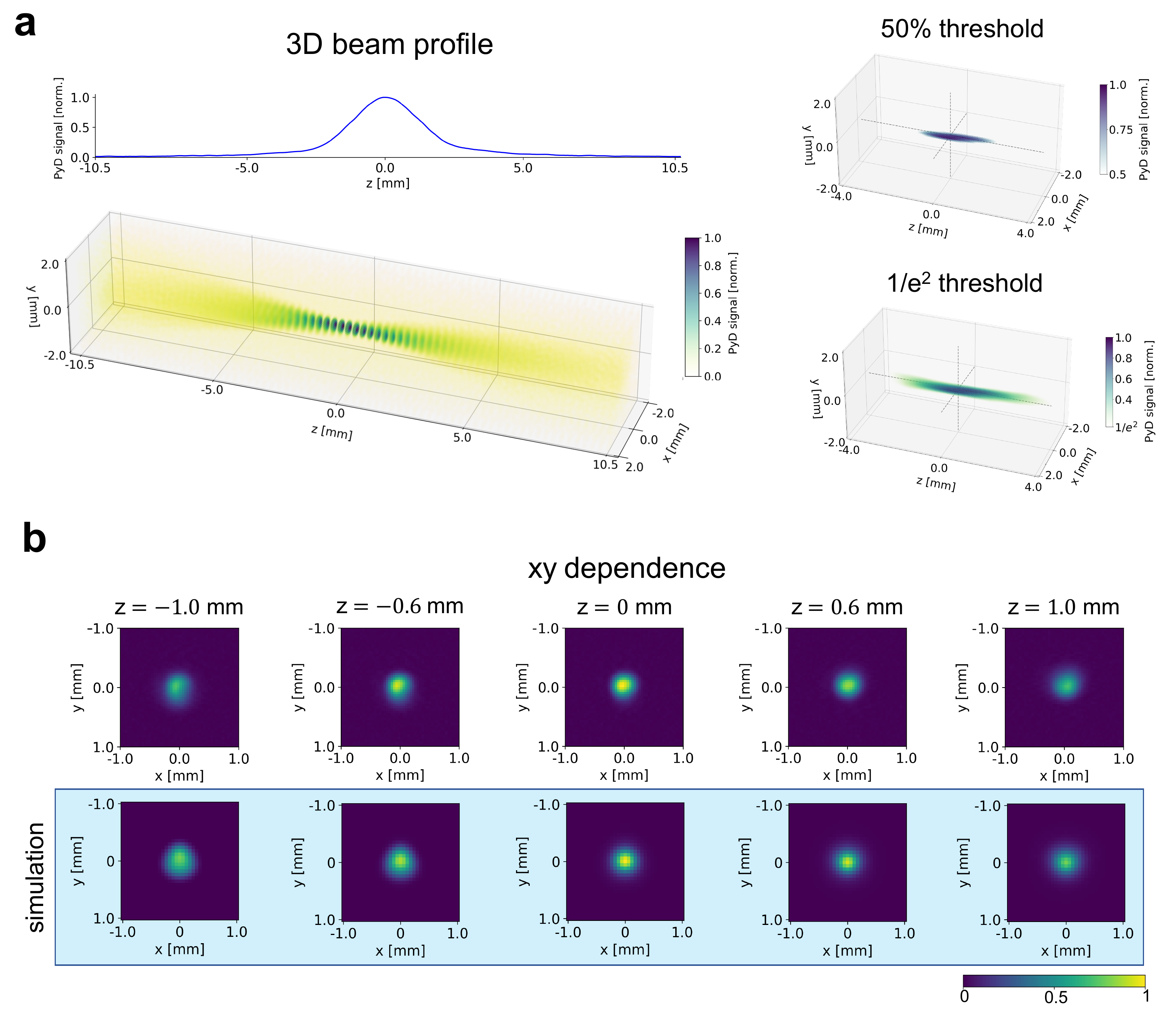}}
\caption{\textbf{Beam focusing onto the imaging plane.} \textbf{a} Measured 3D profile of the beam focused by a 50 mm focal length lens L2 (Tsurupica-RR-CX-1.5-50-SPS, BBLaser Inc., Tokyo, Japan) onto the imaging plane ($z = 0$ mm). The regions of strongest signal (thresholds above 13.5$\%$ ($1/e^2$) and 50$\%$) are subsequently mapped. \textbf{b} Measured and simulated $xy$ beam profiles in the vicinity of the imaging plane.}
\label{fig:focus}
\end{figure}

\begin{figure}[!b]
\makebox[\textwidth][c]{\includegraphics[width=\textwidth]{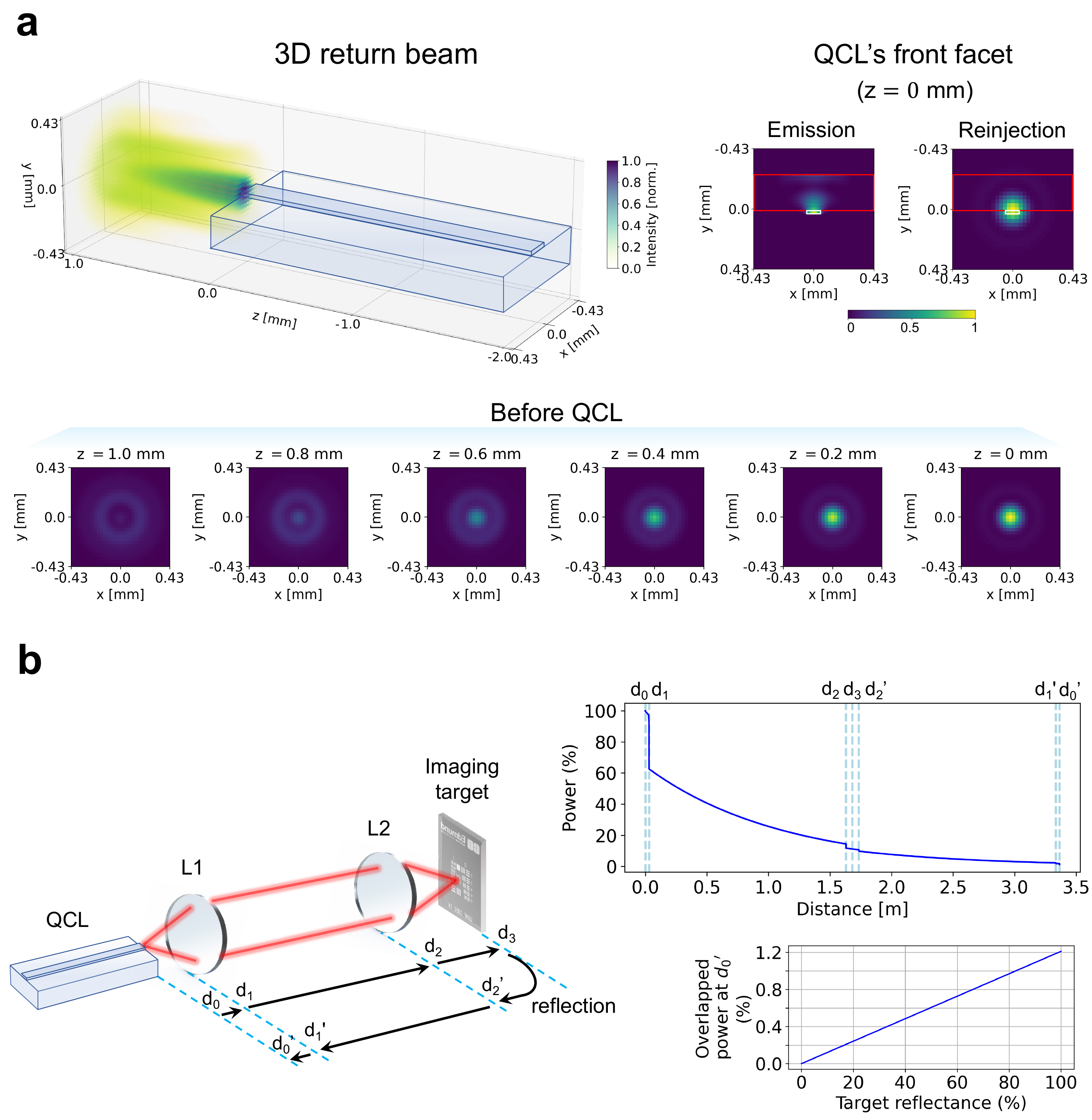}}
\caption{\textbf{Reinjection at the QCL and power loss along the optical system.} \textbf{a} Simulated 3D reinjection profile at the QCL from the reflection of a mirror placed at the focal imaging plane ($z=0$ mm in Fig. \ref{fig:focus}). Emission and reinjection profiles at the QCL's front facet ($z=0$ mm) are displayed, with the white rectangle denoting the QCL's waveguide region and the red rectangle being the QCL's substrate. Several $xy$ beam profiles are shown at different positions along the optical axis ($z > 0$ mm). \textbf{b} Simulated power loss along the optical system. Power percentage relative to distance from the QCL’s front facet is plotted by integrating 2D beam profiles. At $\mathrm{d}_0^{'}$, the power is given by the overlap between emission and reinjection profiles. The change of reinjected power on the imaging target’s reflectance is also computed.}
\label{fig:detection}
\end{figure}

A confocal setup consists of a collimating lens (L1) followed by a focusing lens (L2). The laser exit mirror serves as the confocal aperture$\,$\cite{pawley2006handbook}, as shown in Fig. \ref{fig3}. The THz beam is subsequently reflected, propagates back, and is reinjected into the laser cavity. Both measured and simulated field distributions at carefully selected planes L1, P1, P2, P3, and the imaging plane are shown in Fig. \ref{fig3} and \ref{fig:focus}. The reinjection at the QCL is simulated in Fig.\ref{fig:detection}a. Finally, the power loss along the optical system, in both the forward and return paths, is computed in Fig. \ref{fig:detection}b.

Simulated near-field and corresponding far-field distributions of the QCL are presented in Fig. \ref{fig3}a. The optimal position of the collimating lens ($f = 30$ mm) is found to be $27.8$ mm from the QCL's front facet, as shown in Fig. \ref{fig3}b. The appearance of concentric rings in Fig. \ref{fig3}c is due to diffraction caused by beam clipping at L1, with a clear aperture of $D = 25$ mm, as illustrated in Fig. \ref{fig3}b. This clearly impacts the performance of the system, as we will demonstrate later.

Figure \ref{fig:focus} shows key findings from this study, including the beam profile along the system axis in both forward and return paths. We explored the depth of focus (DOF) and lateral beam extent through simulation and experiment (Fig. \ref{fig:focus}a for 2D and 3D field distributions near the focal point). Figure \ref{fig:focus}b displays the field distributions on both sides of the focal plane (the imaging plane). Beam characterisation is conducted over a $2$ mm axial range near the focal plane, where highly intense radiation ($> 50\%$) is confined. The beam shows a clean circular shape at focus, maintaining its form as it moves away from the focal plane, with some asymmetry due to the QCL's unfiltered emission profile. While further analysis, such as M$^2$ measurements, could be performed to assess beam quality, a qualitative analysis is sufficient for the purposes of this work, which prioritizes PSF measurements.

Finally, reinjection back into the QCL is mapped (Fig. \ref{fig:detection}a), showing optimal reinjection on the front facet of the QCL ($z=0$ mm). The emission and reinjection profiles are compared, revealing that the reinjection extends beyond the waveguide region into the substrate. Power loss analysis (Fig. \ref{fig:detection}b) shows significant losses due to atmospheric absorption, Fresnel reflections, and beam clipping from apertures at L1 ($25$ mm) and L2 ($30$ mm). In backward propagation, atmospheric absorption, target reflectance, and reinjection at the QCL’s front facet contribute more to power loss. Higher reinjected power does not always result in better image quality due to strong feedback that introduces noise and compromises the stability of the system$\,$\cite{qi2021terahertz, ferre2017beam}. To mitigate this, an attenuator (1/4 factor) was used to avoid beam pattern changes and reduce noise, improving system stability. This is an attractive feature towards imaging objects with low reflectivity or when working in an environment prone to scattering and absorption since it induces weak feedback regimes, translating to higher image quality and overall greater system stability.

\FloatBarrier
\clearpage

\subsection*{Comparison between conventional and LFI confocal imaging}\label{sec:spatial_map}
\begin{figure}[!htb]
\makebox[\textwidth][c]{\includegraphics[width=1.\textwidth]{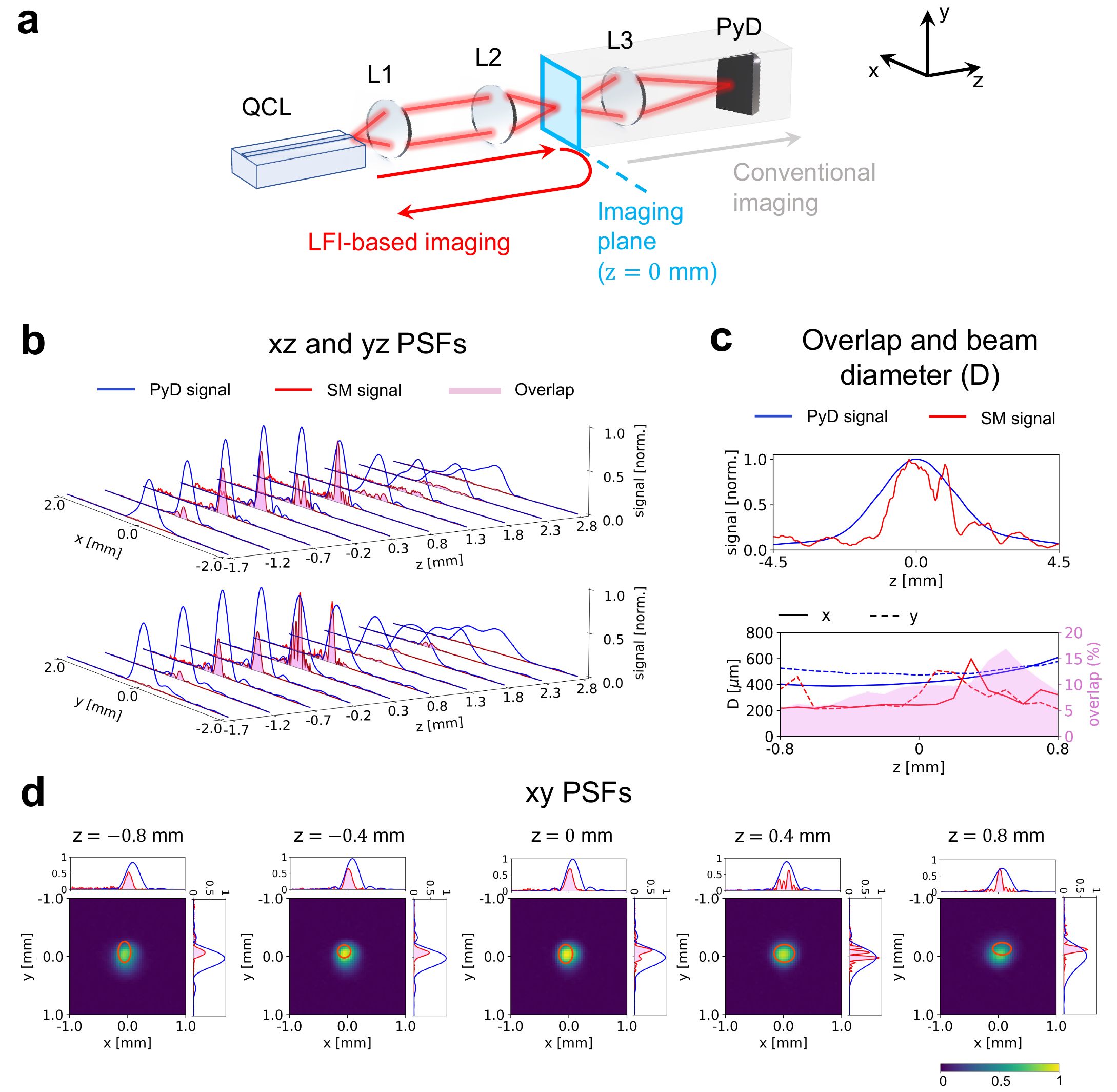}}
\caption{\textbf{Comparison between conventional and LFI imaging PSFs near the imaging plane.} \textbf{a} Optical scheme illustrating LFI and conventional imaging configurations. \textbf{b} 3D PSF of conventional imaging (PyD signal, blue) and LFI imaging (SM signal, red) for planes near the imaging plane ($z=0$ mm). \textbf{c} PSF along the optical axis ($z$) and lateral beam diameters ($D$) are computed to assess axial and lateral resolutions for both systems. \textbf{d} Reinjected beam diameters (red circles) superimposed on the $xy$ focused beam profile (Fig. \ref{fig:focus}b). The overlap percentage is calculated as $100 \times [A_\mathrm{overlap}/(A_\mathrm{PyD} + A_\mathrm{SM} - A_\mathrm{overlap})]$.}
\label{fig_KE_1}
\end{figure}

The 3D PSF, representing the system's impulse response to a point light source, provides insights into the system's resolution properties$\,$\cite{ahi2017mathematical} and is experimentally characterised for both conventional and LFI imaging (Fig. \ref{fig_KE_1}a). Using a knife-edge technique in transmission (PyD signal) and reflection (SM signal) configurations, we obtain the PSF in both imaging setups. Transmission represents conventional imaging, while reflection corresponds to LFI due to reinjected radiation into the laser cavity (see Supplement 1, Section S5). Extending this procedure along the axial direction ($z$) allows the construction of 3D PSF measurements (Fig. \ref{fig_KE_1}b). Closer to the focusing lens, Gaussian profiles are observed, while multi-peak profiles emerge in the LFI PSF for positions farther away ($+z$ direction). The axial PSF and beam diameters are also measured (Fig. \ref{fig_KE_1}c). LFI systems show a narrower full width at half maximum (FWHM) by a factor of 1.5, improving axial resolution in 3D sectioning. Beam diameters in LFI-based systems are up to 2.3 times smaller, enhancing lateral resolution compared to conventional imaging. In regions of high overlap between focused and reinjected profiles ($15\%$ at $z = 0.5$ mm), lateral resolution improves by a factor of 1.6. Fig. \ref{fig_KE_1}d illustrates this axial resolution improvement, with reinjected beam diameters (red line) overlaid on the $xy$ focused beam profiles (Fig. \ref{fig:focus}b). Reinjection from high-intensity regions exhibits astigmatic behavior, leading to a mismatch between $x$ and $y$ beam diameters in axial propagation. To explore the system's potential, imaging of resolution targets and the impact of focusing optics with different numerical apertures will be studied.
\FloatBarrier
\clearpage

\subsection*{The effect of numerical aperture on image resolution}\label{sec:NA}
\begin{figure}[!htb]
\makebox[\textwidth][c]{\includegraphics[width=1.\textwidth]{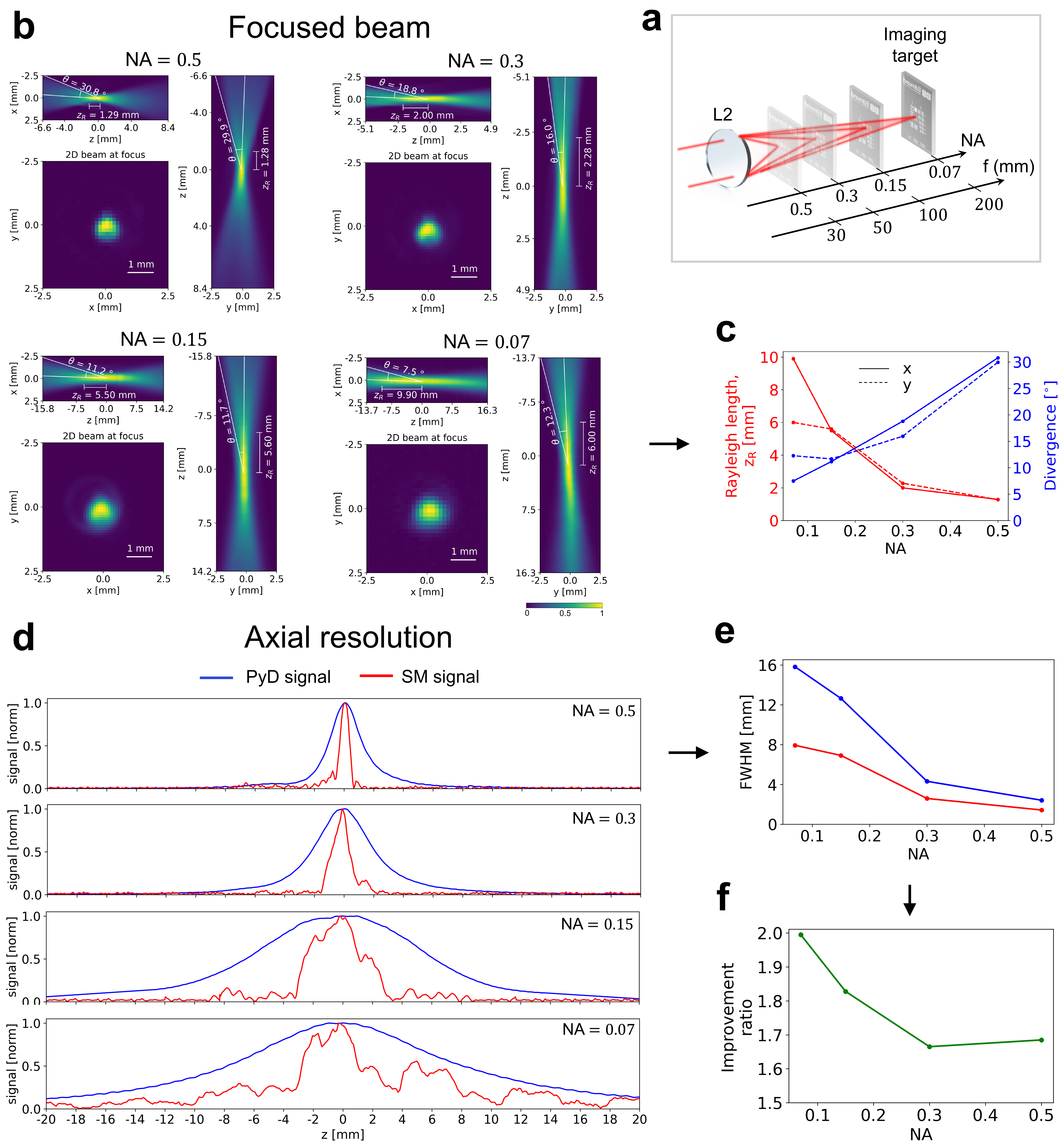}}
\caption{\textbf{Dependence of beam properties and axial resolution on the numerical aperture of the focusing lens L2.} \textbf{a} Different focusing lens apertures are used to inspect the focused beam properties and axial resolution of the system. \textbf{b} Measured 3D focused beam profiles for different NAs. \textbf{c} Rayleigh length and divergence angle of the 3D focused beam profiles for each NA. \textbf{d} PyD and SM signals recorded along the optical axis ($z$) for different NAs. \textbf{e} The FWHM and \textbf{f} the improvement factor ($\mathrm{FWHM}_{\mathrm{PyD}}/\mathrm{FWHM}_{\mathrm{SM}}$) for each NA.}
\label{fig_NA_long}
\end{figure}

The numerical aperture (NA) determines the acceptance angles for target illumination (Fig. \ref{fig_NA_long}a), significantly affecting both lateral and axial resolutions$\,$\cite{pawley2006handbook}. To assess this impact, 3D characterisation of the focused beam is performed (Fig. \ref{fig_NA_long}b). Higher NA leads to increased divergence and a shorter Rayleigh length ($30^{\circ}$, $\mathrm{z}_{\mathrm{R}}=1.3$ mm for $\mathrm{NA}=0.5$), as shown in Fig. \ref{fig_NA_long}c, improving lateral resolution and axial resolution. Higher NA improves axial resolution, with a factor of $1.7$ improvement in LFI configurations ($\mathrm{FWHM}_{\mathrm{PyD}}=2.4$ mm, $\mathrm{FWHM}_{\mathrm{SM}}=1.4$ mm for $\mathrm{NA}=0.5$, Fig. \ref{fig_NA_long}d-f). The lower FWHM results in better axial signal confinement and more detailed sectioning.

\begin{figure}[!b]
\makebox[\textwidth][c]{\includegraphics[width=.9\textwidth]{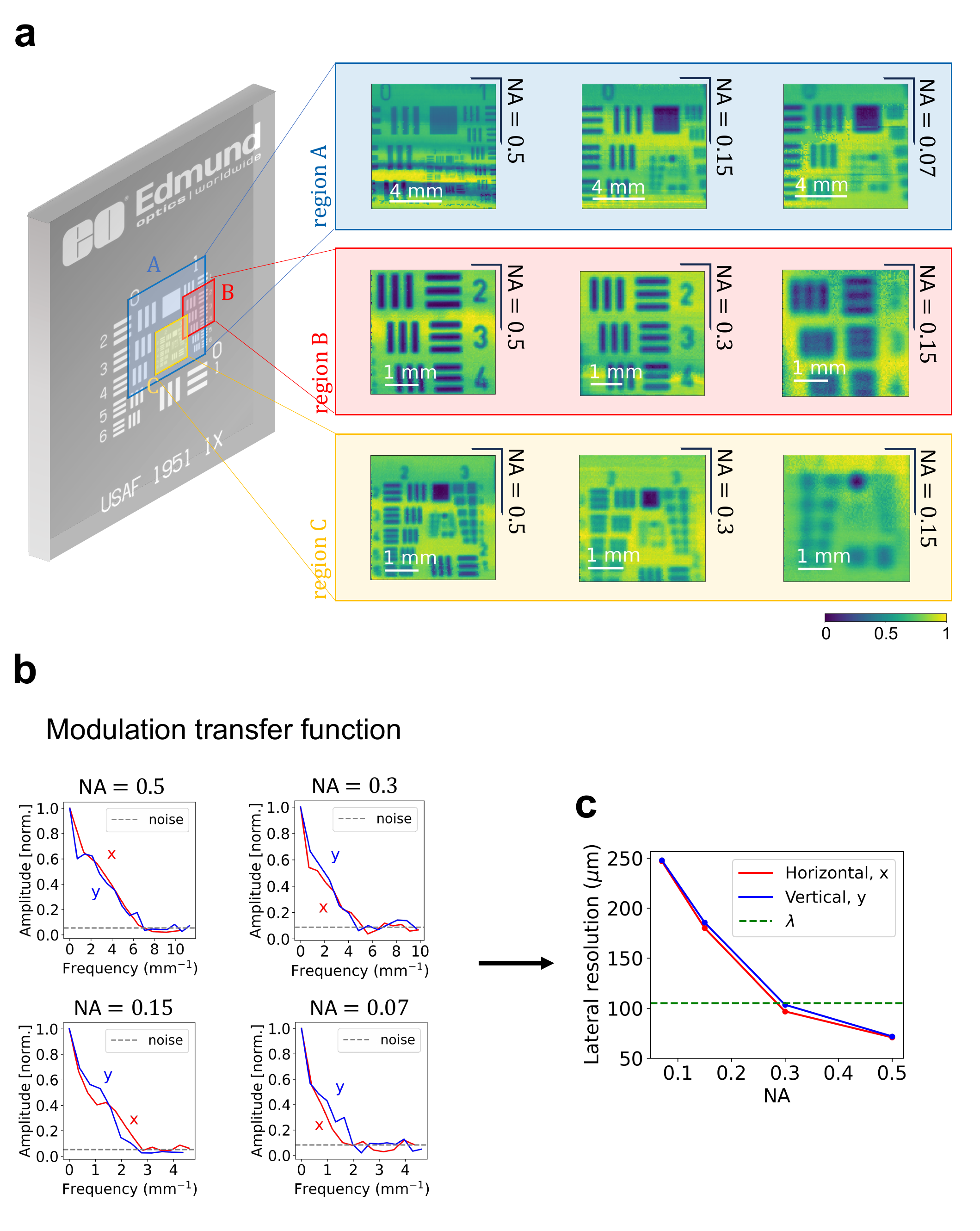}}
\caption{\textbf{Dependence of lateral resolution on the numerical aperture of the focusing lens L2.} \textbf{a} LFI imaging of different areas of a resolution target (2" $\times$ 2" Negative USAF 1951 Hi-Resolution Target, Edmund Optics, NJ, USA) for different focusing NAs. \textbf{b} Computation of the modulation transfer function (MTF) for each NA. \textbf{c} Lateral resolution of the system for each NA.
}
\label{fig_NA_lat}
\end{figure}

Higher NAs further enhance the lateral resolution of LFI imaging systems (Fig. \ref{fig_NA_lat}a). Inspection of subwavelength features of a resolution target is possible via the incorporation of a $0.5$ NA, presenting noticeable improvement upon the $0.3$ NA used when inspecting the system's PSF. The modulation transfer function (MTF) is computed to determine the lateral resolution, achieving values of $71 \mu$m or $0.7 \lambda$ for $\mathrm{NA}=0.5$ (Fig. \ref{fig_NA_lat}b-c). This is attributed to the pinhole effect from the QCL’s waveguide upon reinjection, similar to confocal microscopy techniques$\,$\cite{de2012terahertz}. Interestingly, the astigmatic behaviour shown in Fig. \ref{fig_KE_1} results in distinct $x$ and $y$ lateral resolutions for $\mathrm{NA}=0.3$, while the highest $0.5$ NA displays identical lateral resolutions in both $x$ and $y$, resulting in overall better imaging performance.

Although further resolution improvements are possible with higher NAs and filtering techniques$\,$\cite{de2012terahertz}, this study highlights how NA influences the resolution of LFI-based configurations, which consistently outperform conventional systems both laterally and axially.
\FloatBarrier

\subsection*{Phase-defined depth of focus (DOF)}\label{sec:imaging}
Earlier in the paper, we demonstrated that LFI detection significantly reduces the confocal DOF compared to conventional detection schemes. However, relying solely on amplitude information overlooks one of the key advantages of our coherent imaging approach. 
In this study, we utilise a focusing NA of 0.3 to image a PCB spanning several square centimeters and analyse subwavelength features within a packaged chip. This example also highlights the substantial enhancement in DOF achieved through phase information analysis. By incorporating phase, we reduce the confocal DOF from approximately $10 \lambda$ to $\lambda/5$.  

\begin{figure}[!h]
\makebox[\textwidth][c]{\includegraphics[width=.9\textwidth]{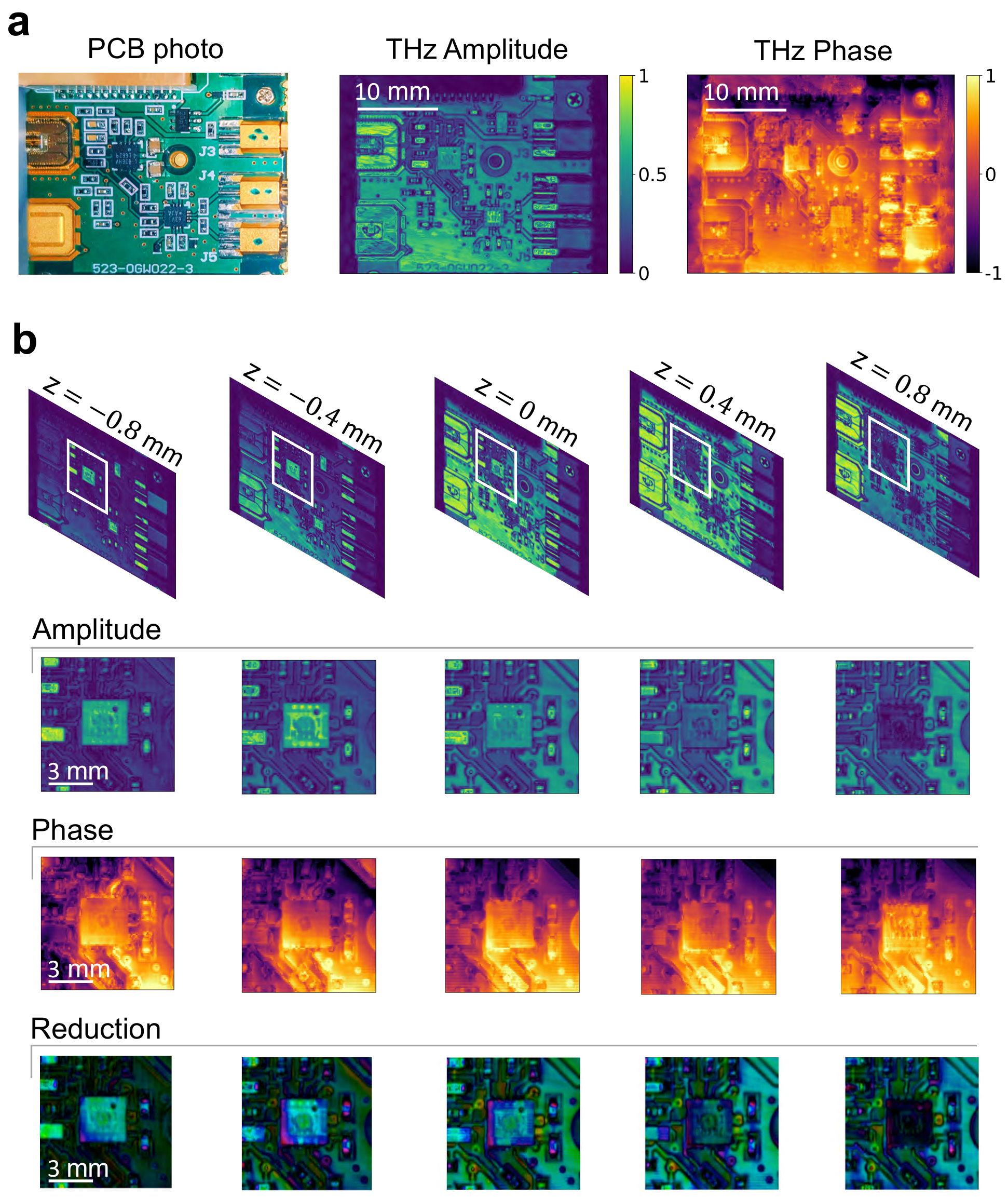}}
\caption{\textbf{LFI tomography imaging of a packaged chip.} \textbf{a} LFI THz amplitude and phase images of a PCB. \textbf{b} LFI THz peak-to-peak tomography imaging of a PCB (first row), LFI THz amplitude (second row) and phase tomography imaging (third row) of one of the PCB's packaged chips (Texas Instrument LMH6629SD). The bottom row consists of reduction images, where amplitude and phase information were combined using the HSV color space (phase for hue, amplitude for value).}
\label{fig_imaging}
\end{figure}

\begin{figure}[!h]
\makebox[\textwidth][c]{\includegraphics[width=.7\textwidth]{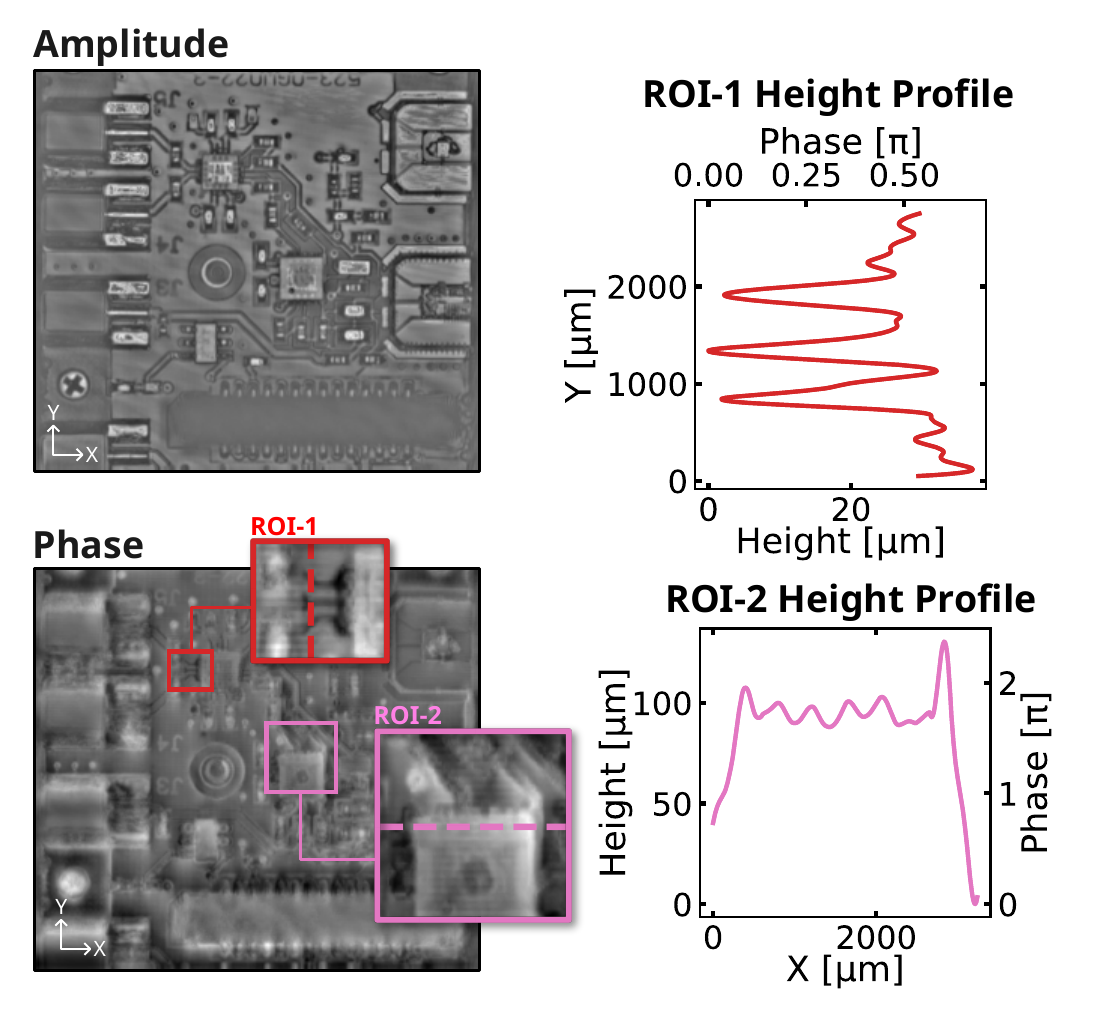}}
\caption{\textbf{Detailed view of the PCB highlighting two regions of interest (ROI-1 and ROI-2).} Depth profiles for ROI-1 (PCB tracks) and ROI-2 (chip package bonding pads) demonstrate structural details and a resolution of $\lambda/5$ from phase contrast converted to equivalent height.}
\label{fig_9}
\end{figure}

Figure \ref{fig_imaging} demonstrates the effectiveness of combining phase and amplitude into a single image to highlight salient features. Figure \ref{fig_9} provides a detailed view of the same PCB, focusing on two regions of interest (ROI-1 and ROI-2). The extracted depth profiles for ROI-1 and ROI-2, corresponding to the PCB tracks and the bonding pads of the chip package, reveal intricate structural details. From the line plots it can be concluded that converting phase contrast information to an equivalent height achieves a resolution of $\lambda/5$.

%
%
%
%
%
%
\FloatBarrier
\clearpage

\section*{Conclusion}
In conclusion, we have demonstrated for the first time that LFI imaging systems provide enhanced resolution over conventional imaging, both laterally and longitudinally. The system's compact design, utilising a single quantum cascade laser for illumination and detection, simplifies alignment by integrating the roles of a detector, confocal pinhole, and illumination source within the laser itself.

Through beam propagation simulations and experimental characterisation, we measured the system's point spread function, revealing significant improvements in lateral and axial resolution through confocal depth-of-focus control. This enhancement is attributed to the LFI confocal aperture, which achieves a DOF as small as $\lambda/5$ by leveraging phase information.

We also examined the influence of the focusing objective's numerical aperture and provided guidelines for optimising the confocal DOF. The system's capabilities were further validated through 3D tomographic imaging, uncovering previously hidden features of a packaged silicon chip with high precision using phase manipulation. A unique advantage over existing methods lies in its ability to achieve high-resolution depth sectioning.

Recent advances in THz imaging and quantum cascade lasers enable compact THz systems for real-time spectral contrast without optical delay lines, external sources, or detectors. Progress in room-temperature QCL operation and frequency combs$\,$\cite{burghoff2014terahertz, lu2019room, khalatpour2021high, silvestri2023frequency} promises portable spectroscopic devices. The system also supports single-pixel imaging techniques, like compressed sensing$\,$\cite{chan2008single, stantchev2020real, dabbicco2021scanless} or diffractive neural networks$\,$\cite{li2023rapid}, eliminating raster scanning and improving acquisition times toward real-time imaging while enabling the \textit{modus operandi} of an actual camera.

\clearpage

\section*{Methods}\label{sec11}

\subsection*{QCL fabrication process}
The THz QCL active region consisting of a 12$\mu$m-thick GaAs/AlGaAs heterostructure was manufactured by molecular beam-epitaxy on a GaAs substrate$\,$\cite{wienold2009low}. Photolithography and chemical etching were then employed to process it into a 150 $\mu$m wide ridge while decreasing the substrate thickness to 200 $\mu$m. Moreover, mechanical cleavage was performed to achieve a physical length of 2 mm.

\subsection*{LFI imaging setup}
Amplitude and phase imaging data were obtained via the fast Fourier transform of the SM waveform. A swept-frequency LFI configuration with pulsed mode operation of the QCL was employed to induce the SM wave$\,$\cite{lim2019coherent}. To enable the swept-frequency technique used for LFI$\,$\cite{rakic2013swept} one has to feed the laser a time-varying drive current, inducing frequency modulation of the laser emission. On such note, the QCL was mounted onto a printed circuit board, and a field programmable gate array (FPGA) controlled laser driver was used to feed a time-varying current through the incorporation of tapered transmission lines. A current pulse with a magnitude of 1.1 A and with a negative ramp of $\approx$ 120 mA was applied over 200 ns with a 10$\%$ duty cycle. In order to read the SM voltage response induced by the ramped current, the measured signal was gated and subsequently amplified by the FPGA controlled laser pulse driver$\,$\cite{lim2019coherent}. To maximise the frequency sweep range of the QCL and prevent ice formation due to atmospheric humidity, the device was kept in a vacuum at 50 K. This was achieved by incorporating a vacuum pump and a Stirling cryocooler with active vibration cancellation. A liquid cooler was placed on the reject side of the cryocooler to further minimise heating of the system. Fast beam steering optics were implemented to conduct 3D imaging. Synchronised scanning of an oscillating mirror with lens L2 enabled access to a larger field of view (20 cm$^2$) while maintaining decent resolution ($\sim$ 100 $\mu$m with NA = 0.3). The system's properties are displayed in Table \ref{tab:imaging_system}. 

\begin{table}[h]
\caption{Imaging system properties.}\label{tab:imaging_system}%
\begin{tabular}{@{}llll@{}}
\toprule
Properties & \\
\midrule
Frequency\footnotemark[1]    & 2.85 THz with 900 MHz frequency sweep  \\
Output power    & 2 mW peak  \\
Image scan area    & 50 (H) $\times$ 40 (V) mm$^2$  \\
Image pixel area    & 1000 (H) $\times$ 400 (V) (0.4 Megapixels)  \\
Pixel size   & 50 (H) $\times$ 100 (V) $\mu$m  \\
Image acquisition time   & 150 seconds (mechanically limited)  \\
\botrule
\end{tabular}
\footnotetext[1]{The fundamental operating frequency was measured using the method presented in ref.$\,$\cite{keeley2017measurement}.}
\end{table}

\subsection*{PSF measurement setup}
A pyroelectric detector (QS1-IL, Gentec-EO, Quebec, Canada) was used to assess the PSF of conventional imaging systems. Due to the slow response times of thermal pyroelectric detectors, modulation of the source was required through the inclusion of an optical chopper (MC1F2, Thorlabs, NJ, USA). A chopper controller (MC2000, Thorlabs, NJ, USA) set the modulation frequency ($f_\mathrm{m} = 100$ Hz) which was used by a lock-in amplifier (7280 DSP, AMETEK Inc., TN, USA) to increase the sensitivity of the pyroelectric detector and decrease the measured noise. For practical reasons the setup was automated using Python 3.9. On such note, most of the experimental setup was controlled by a computer: the cryocooler, the laser driver which enabled the emission of the QCL as well as the acquisition of the SM signal, the lock-in detection system for acquisition of the pyroelectric signal, and the mechanical translation stages for scanning automation.

\subsection*{Beam propagation and power loss}
The beam propagation software was developed based on a fundamental understanding of free space wave propagation. Propagation of an optical field in free space can be described as a linear system between an input complex field located at a plane $z=0$ and an output complex field located at a distance $z=d$$\,$\cite{novotny2012principles}:
\begin{equation}
\tilde{\mathbf{E}}(k_x,k_y,d) = \tilde{\mathbf{H}}(k_x,k_y,d) \cdot \tilde{\mathbf{E}}(k_x,k_y,0),
\label{eq:k_propagation}
\end{equation}
where $k_x$ and $k_y$ are the spatial frequencies along $x$ and $y$ directions, respectively, and $\tilde{\mathbf{H}}$ represents the propagator in reciprocal space:
\begin{equation}
\tilde{\mathbf{H}}(k_x,k_y,d)=\mathrm{e}^{\pm ik_zd},
\end{equation}
also known as the optical transfer function. The sign $\pm$ accounts for the direction of propagation along $z$ (forward and backward) and $k_z$ denotes the propagating wave vector:
\begin{equation}
k_z = \sqrt{k_0^2 - k_x^2 - k_y^2},
\end{equation}
where $k_0 = 2 \pi/ \lambda_0$, with $\lambda_0$ being the laser operating wavelength ($\lambda_0 = 105$ $\mu\mathrm{m}$ or $f = 2.85$ THz).
Therefore, given an input optical field $\tilde{\mathbf{E}}(x,y,z=0)$ in real space, we can obtain the output propagated field by computing the inverse Fourier transform of Eq. (\ref{eq:k_propagation}):
\begin{equation}
\tilde{\mathbf{E}}(x,y,d)=\mathrm{F}^{-1} \left\{ \tilde{\mathbf{E}}(k_x,k_y,d) \right\} = \mathrm{F}^{-1} \left\{ \tilde{\mathbf{H}}(k_x,k_y,d) \cdot \mathrm{F} \left\{ \tilde{\mathbf{E}}(x,y,0) \right\} \right\},
\end{equation}
where $\tilde{\mathbf{E}}(x,y,0) = E_0(x,y)\mathrm{e}^{i\phi(x,y)}$, with $E_0(x,y)$ representing the field amplitude and $\phi(x,y)$ its phase. Computationally, this translates to the implementation of fast Fourier transforms on an $xy$ matrix of $m\times p$ complex field values, with $m$ and $p$ being the number of rows and columns, respectively.

In order to accurately simulate the propagation through the optical system one has to account for optical components such as lenses and mirrors$\,$\cite{saleh2019fundamentals}. The transmission of a complex field $\tilde{\mathbf{E}}(x,y)$ through a lens introduces a quadratic phase shift:
\begin{equation}
\tilde{\mathbf{E}}_{\mathrm{t}}(x,y) = \tilde{\mathbf{E}}(x,y) \mathrm{e}^{\left[ ik_0 \frac{x^2+y^2}{2f} \right]},
\label{eq:quadratic_phase_shift}
\end{equation}
where $\tilde{\mathbf{E}}_{\mathrm{t}}(x,y)$ is the transmitted complex field and $f$ is the focal length of the lens. On the other hand, reflection from a planar surface (eg. mirror or imaging target) introduces a phase shift of $\pi$ to the incident field:
\begin{equation}
\tilde{\mathbf{E}}_{\mathrm{r}}(x,y) = \tilde{\mathbf{E}}(x,y)\mathrm{e}^{i\pi},
\label{eq:phase_match_mirror}
\end{equation}
where $\tilde{\mathbf{E}}_{\mathrm{r}}(x,y)$ is the reflected complex field. The loss of power across each optical component is obtained by computing the power relative to the corresponding 2D field profile: 
\begin{equation}
P = \int_{\mathrm{S}} \mathbf{I} \cdot \mathrm{dS} = \frac{c\epsilon_0n}{2}\int_x \int_y |\tilde{\mathbf{E}}(x,y)|^2 \mathrm{d}x\mathrm{d}y, 
\label{eq:power_loss}
\end{equation}
where $\mathbf{I}$ is the field intensity, $\mathrm{S}$ is the surface area of the 2D field profile, $c$ is the speed of light, $\epsilon_0$ the vacuum permittivity and $n$ the refractive index of the medium. Power loss due to reflections at the different interfaces of the optical components is accounted for by computing the Fresnel reflectance $R$ upon normal incidence:
\begin{equation}
P_t = (1-R) P_i, \quad R = \left| \frac{n_1-n_2}{n_1+n_2} \right|^2,
\label{eq:R_fresnel}
\end{equation}
where $P_t$ is the transmitted power, $P_i$ is the incident power, and $n_1$ and $n_2$ represent the refractive index of the incident and transmitted media, respectively. Moreover, one should also account for air humidity absorption and partial reinjection at the QCL's front facet. The presence of air humidity results in an exponential decrease of the transmitted power$\,$\cite{slocum2013atmospheric}: 
\begin{equation}
P_t = P_0 \mathrm{e}^{-\alpha l},
\end{equation}
where $P_0$ is the initial power, $\alpha$ is the absorption coefficient and $l$ the path length.

\section*{Data availability}
All raw data that supports the findings of this study is openly available at \url{https://doi.org/10.48610/bf3e83f}. 

\section*{Code availability}
The code used to simulate the beam propagation is openly available on GitHub: \url{https://github.com/jrgsilv/beam-propagation}. The code used for analysing the experimental data is available from the corresponding author upon request. 
\clearpage

\backmatter
\bibliography{sn-bibliography}

\clearpage 

\section*{Acknowledgments}
This work has been supported by the funding from Australian Research Council (DE170100241, DP200101948, DP210103342); Engineering and Physical Sciences Research Council (EP/J002356/1, EP/P021859/1); UK Research and Innovation (Future Leader Fellowship MR/S016929/1); Advance Queensland Industry Research Fellowships program and UQ Amplify Fellowship.
The authors acknowledge Marcos Maestre Morote, Mickael Mounaix, Andrew Komonen, Joel Carpenter, Mayuri Kashyap and Lucas Cardoso for the helpful discussions. The authors are also thankful for the valuable contribution from the Stack Overflow community to numerous coding issues.

\section*{Author contributions}
J.S.: writing – original draft (lead); investigation (lead); formal analysis (lead); visualisation (lead); methodology (equal); conceptualisation (supporting); software (supporting).
M.P.: software (lead); methodology (equal); writing — reviewing and editing (equal); visualisation (supporting); conceptualisation (supporting); formal analysis (supporting).
K.B.: methodology (equal); writing — reviewing and editing (equal); supervision (equal); resources (equal); investigation (supporting); conceptualisation (supporting); visualisation (supporting); formal analysis (supporting). 
M.G.: writing — reviewing and editing (equal); investigation (supporting); formal analysis (supporting); visualisation (supporting).
T.G.: methodology (equal); resources (equal). 
J.T.: methodology (equal); writing — reviewing and editing (equal); software (supporting); formal analysis (supporting).
J.H.: resources (equal); software (supporting).
Y.L.L.: methodology (equal); resources (equal); conceptualisation (supporting).
T.T.: methodology (equal); conceptualisation (supporting).
X.Q.: supervision (equal); writing — reviewing and editing (equal). 
B.C.D.: visualisation (supporting).
T.Z.: software (supporting); visualisation (supporting).
H.S.L.: methodology (equal); software (supporting).
D.I.: methodology (equal); resources (equal).
Y.H.: methodology (equal); resources (equal).
L.L.: methodology (equal); resources (equal).
A.V.: methodology (equal); software (supporting).
E.H.L.: methodology (equal); resources (equal).
A.G.D.: methodology (equal); resources (equal).
P.D.: methodology (equal); resources (equal).
A.D.R: conceptualisation (lead); funding acquisition (lead); methodology (equal); supervision (equal); writing — reviewing and editing (equal); investigation (supporting); visualisation (supporting).


\section*{Competing interests}
The authors declare no competing interests.


\end{document}